\documentclass{aa}  
\usepackage{graphicx}
\usepackage{txfonts}
\begin{document}

   \title{Feeding the spider with carbon}

   \subtitle{[CII] emission from the circum galactic medium and active galactic nucleus}

   \author{C. De Breuck
          \inst{1}
          \and
          A. Lundgren\inst{1}
          \and
          B. Emonts\inst{2}
          \and
          S. Kolwa\inst{3,4}
          \and
          H. Dannerbauer\inst{5,6}
          \and
          M. Lehnert\inst{7}
          }

     \institute{European Southern Observatory,
              Karl Schwarzschild Stra\ss e 2, 85748 Garching, Germany \email{cdebreuc@eso.org}\label{inst1}
              \and
              National Radio Astronomy Observatory, 520 Edgemont Road, Charlottesville, VA 22903, USA\label{inst2}
              \and
              Inter-University Institute for Data Intensive Astronomy, Department of Astronomy, University of Cape Town, Rondebosch 7701, South Africa\label{inst3}
              \and
               Physics Department, University of Johannesburg, 5 Kingsway Ave, Rossmore, Johannesburg, 2092, South Africa\label{inst4}
               \and 
               Instituto de Astrof\'\i sica de Canarias (IAC), E-38205 La Laguna, Tenerife, Spain\label{inst5}
               \and
               Universidad de La Laguna, Dpto. Astrof\'\i sica, E-38206 La Laguna, Tenerife, Spain \label{inst6}
              \and
              Universit\'e Lyon1, ENS-Lyon, CNRS, Centre de Recherche Astrophysique de Lyon UMR5574, F-69230, Saint-Genis-Laval, France\label{inst7}
 }

   \date{Received 2021 July 23; accepted 2021 December 22}

% \abstract{}{}{}{}{} 
% 5 {} token are mandatory
 
  \abstract{
We present the detection of [CII]\,158\,$\mu$m emission from the Spiderweb galaxy at $z$=2.1612 using the Atacama Pathfinder EXperiment. The line profile splits into an active galactic nucleus (AGN) and circum galacic medium (CGM) component previously identified in CO and [CI]. We find that these individual [CII] components are consistent in terms of CO and far-IR luminosity ratios with the populations of other $z$$\gtrsim$1 AGN and dusty star-forming galaxies. The CGM component dominates the [CII] emission in the 10\arcsec\ APEX beam. Although we do not have spatially resolved data, the close correspondence of the velocity profile with the CO(1-0) detected only on scales of tens of kiloparsecs in CO(1-0) suggests that the [CII] emission is similarly extended, reminiscent of [CII] halos recently found around $z$$>$5 galaxies. Comparing the first four ionization states of carbon, we find that the atomic [CI] emission is dominant, which increases its reliability as a molecular mass tracer. Our [CII] detection at 601.8\,GHz also demonstrates the feasibility to extend the frequency range of ALMA Band 9 beyond the original specifications. }

  \maketitle
%
%-------------------------------------------------------------------

\section{Introduction}
The important role of the circum galactic medium (CGM) as a reservoir feeding star forming gas to galaxies is now well established \citep[e.g.][]{dekel2009}. However, especially at high redshift, our knowledge of the CGM is still mostly limited to the brightest emission lines (e.g. Lyman-$\alpha$), which mainly trace the warm gas and have the disadvantage of being poor tracers of the intrinsic velocity of the gas due to resonant scattering effects, and being ionized by a range of physical processes involving an active galactic nucleus (AGN), star formation, and shocks from inflows or outflows \citep{tumlinson2017,daddi2021}.

A more direct way to study the link between the CGM and star formation is to observe cold gas containing molecular hydrogen, which is the fuel for forming stars. This cold gas can be detected in the (sub)millimetre using bright CO \citep[e.g.][]{cicone2014,emonts2016,ginolfi2017,li2021} or fine structure lines \citep[e.g.][]{cicone2015,fujimoto2020,herrera-camus2021}. Most of these results, especially those using the [CII]\,158\,$\mu$m line (hereinafter [CII]), appear to be tracing AGN or star formation driven outflows rather than an extended gas reservoir feeding the central galaxy. Most importantly, by its large spatial scale nature, any interferometer over-resolves a significant part of the extended CGM emission, in particular at the high observing frequencies of [CII] \citep{carniani2020,novak2020,decarli2021}. This is where sensitive single-dish submillimetre (submm) telescopes can play an important role. Due to their limited collecting area, we can currently only target the brightest emission lines such as [CII].

In this Letter, we present Atacama Pathfinder EXperiment (APEX) [CII] observations of the Spiderweb galaxy at $z$=2.1612, one of the best studied high redshift radio galaxies (HzRG), located at the centre of a protocluster \citep[e.g.][]{pentericci2000,miley2006}. \citet{emonts2013} first detected CO(1-0) in the Spiderweb galaxy using the Australia Telescope Compact Array (ATCA). Deeper observations showed that this emission splits into two components dominated by the AGN and the CGM, which is over-resolved in longer baseline observations with the Karl J. Jansky Very Large Array (VLA) \citep{emonts2016}. The extended CGM emission follows the diffuse UV light from young stars found with the {\it Hubble Space Telescope} \citep{hatch2008}. While the CO(1-0) line traces the cold molecular gas, it is unfortunately rather faint. This is where the bright [CII] line presents a good alternative. However, at $z$=2.1612, the [CII] line falls at 601.8\,GHz, just below the edge of the ALMA Band 9 receivers, designed to cover 602 to 720\,GHz \citep{baryshev2015}. The upgraded version of this receiver installed in the Swedish ESO PI for APEX \citep[SEPIA;][]{belitsky2018} has extended the frequency range to 578--738\,GHz, which now allows one to observe the [CII] line in the Spiderweb galaxy.

Throughout this paper, we assume a $\Lambda$CDM cosmology with $H_0$=67.8\,km\,s$^{-1}$Mpc$^{-1}$, $\Omega_m$=0.308, and $\Omega_{\Lambda}$=0.692 \citep{planck2016}. At $z$=2.1612, this corresponds to a luminosity distance $D_L$=17.5\,Gpc and a scale of 8.5\,kpc/\arcsec.

%--------------------------------------------------------------------
\section{Observations and data reduction}

\begin{figure}
   \centering
   \includegraphics[width=0.45\textwidth,angle=0]{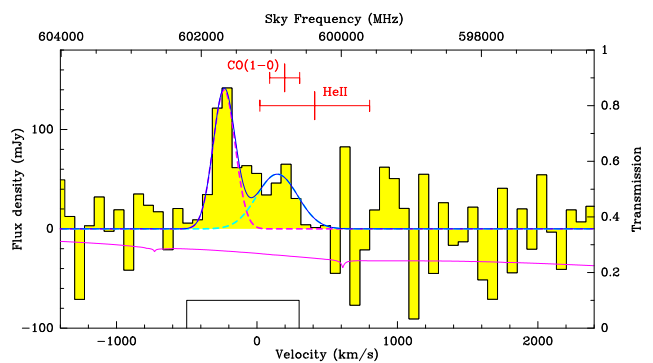}
   \caption{APEX/SEPIA660 [CII] spectrum at a velocity resolution of 70 km/s. The solid magenta line shows the average transmission during the observations (assuming elevation=60$^{\circ}$, PWV=0.4 mm, and ambient temperature=0$^{\circ}$ Celsius). We marked the redshifts and widths of the HeII line \citep{silva2018} and the AGN component in the CO(1-0) line \citep{emonts2016}. The magenta- and cyan-dashed lines show the double Gaussian fit (sum in blue), where the redshift of the higher velocity component has been fixed to that of CO(1-0). The black line is the velocity exclusion region used for the baseline subtraction.}
              \label{fig:CII}%
    \end{figure}

We observed the [CII]\,158\,$\mu$m ($\nu_{\rm rest}$=1900.539 GHz) using the Swedish-ESO PI Instrument for APEX \citep[SEPIA;][]{belitsky2018} on the APEX telescope \citep{guesten2006}. The data were obtained under ESO project E-0106.A-1003A-2020 during five nights in December 2020. The total on-source integration time centred on RA=11$^{\rm h}$40$^{\rm m}$48\fs 34 DEC=$-$26$^{\circ}$29$^{\prime}$08\farcs 66 was 4.3\,h and the telescope time including all overheads and calibrations was 17.1\,h. The precipitable water vapour (PWV) was in the range 0.3--0.6\,mm, corresponding to a transmission 0.21 to 0.45 at the science frequency. While the CO redshift of 2.1612 used by \citet{emonts2018} places the [CII] line at 601.204\,GHz, we preferred to tune the receiver to 602.26\,GHz in the lower sideband to centre the line in the middle of one of the two 4\,GHz wide backend units, so as to avoid any edge effects affecting the line profile in the small overlap region between the two backends.
%Given the (then) assumed redshift z=2.16, the emission line was expected at 601.44 GHz, but for technical reasons the receiver was tuned to 602.44 GHz (in the lower sideband). 
We used the wobbler in symmetrical mode with an amplitude of 20$\arcsec$ and frequency of 1.5 Hz. Pointing and calibration was checked regularly against V Hya and IRC+10216 using the CO(J=6-5) emission line. We estimated the overall calibration uncertainty at 15\% and that the pointing accuracy was typically within 2$\arcsec$.  The data were reduced using the standard procedures in the Continuum and Line Analysis Single-dish Software \citep[CLASS;][]{pety2005}. We aligned both backend units by a simple average in the overlap region and fitted a first order baseline to each scan before averaging them\footnote{We also performed baseline subtraction on individual backends and then combined the data, providing consistent results.}. In order to avoid subtracting the broad wings of the emission line, we excluded a region $\pm$500\,km\,$^{-1}$ from the expected line centre (see in Fig.~\ref{fig:CII}).

Main beam characteristics have been determined from de-convolved continuum slews across Mars in September 2020. At 602\,GHz, this yields a mean beam size of $\theta_{mb}$ = 9\farcs 8$\pm$0\farcs 1, which we confirmed to be consistent with cross scans on the pointing sources. To determine the main beam efficiency, we used all Mars cross scans\footnote{See http://www.apex-telescope.org/telescope/efficiency/index.php} obtained between September and December 2020 to obtain a reliable fit of the Ruze formula (J.-P. P\'erez-Beaupuits, private communication). This yields a main beam efficiency $\eta_{mb}$ = 0.58$\pm$0.04 and an antenna gain of Jy/K=48.5$\pm$4.
%Converting the results to our science frequency they become main beam size $\theta_{mb}$ = 9.8$\pm$0.1$\arcsec$, antenna gain of Jy/K=47 $\pm$ 4 and main beam efficiency $\eta_{mb}$ = 0.56$\pm$ 0.03.
Fig.\,\ref{fig:CII} shows the [CII] spectrum binned to 70\,km\,s$^{-1}$ with the atmospheric transmission \citep{pardo2001} overplotted to illustrate the smooth gradient over the spectral range.

\section{Results}
\begin{table}[ht]
\caption{Observational parameters of the Spiderweb galaxy} % title of Table
\label{table:obsparams}      % is used to refer this table in the text
\begin{center}                          % used for centering table
\begin{small}
\begin{tabular}{l c c c c c c}        % centered columns (4 columns)
\hline\hline                 % inserts double horizontal lines
Component & Velocity offset$^*$ & $I_{[CII]}$ & FWHM & $L_{[CII]}$ \\
 & km\,s$^{-1}$ & Jy\,km\,s$^{-1}$ & km\,s$^{-1}$ & $10^9 L_{\odot}$ \\
\hline
%CGM & -245$\pm$19 & 18$\pm$10 & 112$\pm$92  & 3.4$\pm$1.9 \\
%AGN & -11$\pm$145 & 23$\pm$10 & 610$\pm$240 & 4.4$\pm$1.9 \\
CGM & -234$\pm$25 & 28$\pm$9 & 185$\pm$85  & 5.3$\pm$1.7 \\
AGN & 144.6$^{\dagger}$ & 20$\pm$11 & 340$\pm$190 & 3.8$\pm$2.1 \\
\hline                                   %inserts single line
\end{tabular}
\end{small}
\end{center}
$^{*}$ Relative to z=2.1612.\\
$^{\dagger}$ Fixed to CO(1-0) redshift \citep{emonts2018}.
\end{table}

%We found that the the velocity integrated intensity for the [CII] was 0.84 $\pm$ 0.22 K km s$^{-1}$ ($T_A^*$ scale) in the range -500 to 500 km/s, where the velocity 0 km/s corresponds to the sky frequency of 601.20407 GHz (corresponding to z=2.16121763). There are apparently two main velocity components; a narrow at -243 km/s with a width of 128 km/s and a broader at +56 km/s with a width of 436 km/s. The flux of the two components are 0.37 and 0.47 K km s$^{-1}$ ($T_A^*$ scale), respectively, and the error bars are in both cases of the order of 0.2 K km s$^{-1}$. 

We detect the [CII] line with a velocity integrated intensity of 48$\pm$11\,Jy\,km\,s$^{-1}$ in the range -500 to 500 km\,s$^{-1}$, where 0 km\,s$^{-1}$ corresponds to the sky frequency of 601.204\,GHz \citep[corresponding to $z$=2.1612;][]{emonts2018}. The spectral profile clearly deviates from a single Gaussian, and it consists of two main velocity components listed in Table~\ref{table:obsparams}.
%: a narrow at -245$\pm$19\,km\,s$^{-1}$ with a width of 112$\pm$92\,km\,s$^{-1}$ and a broader at -11$\pm$145\,km\,s$^{-1}$ with a width of 610$\pm$240\,km\,s$^{-1}$. The integrated line fluxes of the two components are 18$\pm$10 and 23$\pm$10\,Jy\,km s$^{-1}$, respectively.

The [CII] spectral profile reflects the different velocity components covered by the 10\arcsec\ APEX beam. The CO and [CI] detected galaxies of \cite{emonts2018} with velocities within the observed [CII] line are located 9\arcsec\ to 22\arcsec from the APEX pointing, which is well outside of the APEX beam. Other  companion galaxies are located within the APEX beam, but none of them have velocities within the [CII] profile \citep{pentericci2000,kurk2004,kuiper2011}. One exception could be a very tentative ($<$2$\sigma$) detection at 599.3\,GHz, which is close to the expected $z$=2.1701$\pm$0.0016 of source \#5 of \citet{kuiper2011}, located at the eastern edge of the host galaxy.

The AGN-dominated region\footnote{We refer to the region within $\sim$6\,kpc from the radio core as AGN. This includes (most of) the AGN host galaxy.} has been detected in a range of emission lines. In the rest-frame UV, the narrow-line region has a velocity width of $\sim$2000\,km\,s$^{-1}$ \citep{silva2018}, while the H$\alpha$ line has a width of 15000\,km\,s$^{-1}$, which can only originate from the AGN broad-line region \citep{nesvadba2006,nesvadba2011,humphrey2008}. The non-resonant HeII\,$\lambda$1640\AA\ recombination line is commonly used as the best tracer of the AGN systemic redshift $z_{\rm HeII}$=2.1623$\pm$0.0011 \citep{silva2018}. This corresponds within the uncertainties with the $z_{\rm CO,VLA}=2.1617\pm0.0003$ of the AGN component in the CO(1-0) line identified by \citet{emonts2016}. As the HeII line has a 3$\times$ higher FWHM than the CO(1-0) (Fig.~\ref{fig:CII}), the latter can provide a more accurate redshift constraint, despite the limited signal-to-noise of the CO(1-0) data. In addition, as the [CII] predominantly traces the photo-dissociation regions, it is expected to originate from the same gas as the CO(1-0), while the HeII is tracing the more extended photo-ionized gas. We thus assume the CO(1-0) AGN redshift and the nominal redshift of the AGN.

\begin{figure*}[ht]
   \centering
   \includegraphics[width=14cm]{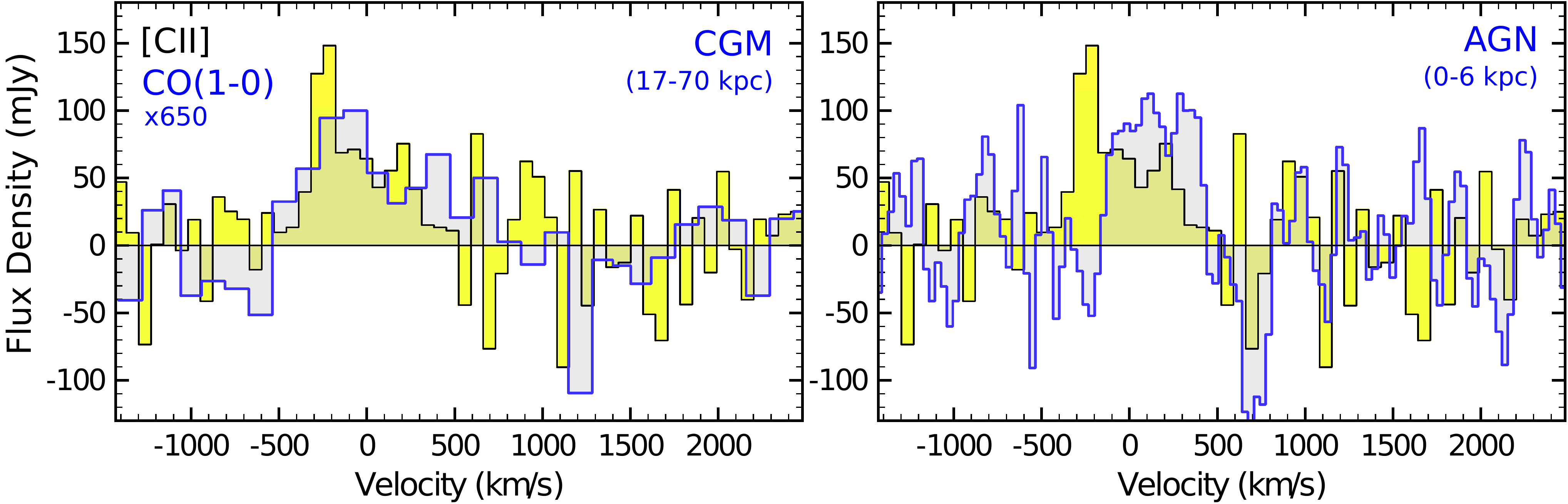}
   \caption{Total [CII] spectrum (yellow histogram) with CO(1-0) from different regions overlaid in blue.  In both panels, the CO(1-0) was scaled up in flux by a factor of 650 for easy comparison with [CII].  {\it Left:} CO(1-0) emission on scales of 17-70 kpc in the CGM around the Spiderweb galaxy. The CO(1-0) spectrum was made from low-resolution ATCA data with the central emission subtracted \citep[see][for details]{emonts2018}. {\it Right:} CO(1-0) emission in the inner $\sim$6 kpc of the radio galaxy, from high-resolution VLA data \citep{emonts2016}. The CO(1-0) emission on $\sim$6-17 kpc scales around the AGN is not captured reliably in this figure due to the difference in resolution between the ATCA and VLA data.}
              \label{fig:CO_CII}%
    \end{figure*}

The molecular and atomic gas in the Spiderweb galaxy show an even more complex structure. \citet{gullberg2016} found at least three spatially and spectrally separated components detected in the [CI]\,370\,$\mu$m ($^3$P$_2$$-$$^3$P$_1$) line. Two of these components with velocity widths of 270 and 1100\,km\,s$^{-1}$ appear to be associated with the AGN, with an additional 230\,km\,s$^{-1}$ wide component offset by +360\,km\,s$^{-1}$. \citet{emonts2018} found a very similar total profile in the [CI]\,609\,$\mu$m ($^3$P$_1$$-$$^3$P$_0$) and CO(4-3) lines. On more extended CGM scales (17--70\,kpc), \citet{emonts2016,emonts2018} found that the CO(1-0) is blueshifted by a few hundred km\,s$^{-1}$ with respect to the CO(1-0) in the AGN. Fig.~\ref{fig:CO_CII} compares these AGN and CGM components of CO(1-0) with our [CII] velocity profile, and it illustrates a striking correspondance. We therefore interpret the narrow [CII] component at -234\,km\,s$^{-1}$ to originate from the CGM predominantly.

%-------------------------------------- Two column figure (place early!)
\section{Discussion}

Our [CII] detection is not only the first one reported in a HzRG, but also one of the few reported in the CGM of a massive high-redshift galaxy \citep[e.g.][]{cicone2015}. This allows us to compare the Spiderweb galaxy with other high-$z$ galaxies, and to provide new insight into the physical conditions in the CGM.

%L_AGN=17500^2*1.04e-3*23*1900.54/3.1612=4.4e9
% Our detection corresponds to $L_{\rm [CII],AGN}$=(4.4$\pm$1.9)$\times$$10^9$$L_{\rm \odot}$. Similarly, $L_{\rm [CII],CGM}$=(3.4$\pm$1.9)$\times$$10^9$$L_{\rm \odot}$. 
We first compared the total [CII] luminosity of the Spiderweb galaxy with other sources. A key property is the FIR luminosity, integrated over the source based on {\it Herschel} observations. \citet{seymour2012} report an IR (8--1000\,$\mu$m) $L_{\rm IR}$=2$\pm$0.3$\times10^{13}\,L_{\odot}$, splitting in 1.2 and 0.8$\times$10$^{13}$$L_{\odot}$ for the AGN and starburst (= CGM) components, respectively. To convert from IR to FIR (42--500\,$\mu$m), we assumed $L_{\rm FIR}=0.5 \times L_{\rm IR}$; this implies a $L_{\rm [CII]}/L_{\rm FIR} \approx 9\times10^{-4}$, which is close to the average for dusty star-forming galaxies \citep[DSFG;][]{gullberg2015}, but lower than for $z$$\sim$2 main sequence galaxies \citep{zanella2018}. Separating both the dust and [CII] emission into AGN and CGM-dominated components, the ratios become $6\times10^{-4}$ for the AGN and $1.3\times10^{-3}$ for CGM component. These different $L_{\rm [CII]}/L_{\rm FIR}$  values are consistent with those for other high-$z$ AGN and star-forming galaxies reported by \citet{gullberg2015}.
%The [CII]/FIR luminosity in the Spiderweb galaxy (a bona fide type 2 AGN) is thus similar to that seen in high-redshift quasars, which is expected for isotropic and (mostly) optically thin tracers. 
We also note that with a 0.09\% contribution to the $L_{\rm FIR}$, the [CII] line unlikely affects the {\it Herschel} 500\,$\mu$m photometry \citep{smail2011,seymour2012}. Interestingly, the star formation rate SFR=1400$\pm$150\,$M_{\odot}$\,yr$^{-1}$ \citep[where the AGN component has been spectrally removed in the SED;][]{seymour2012} is exactly on the SFR-$L_{\rm [CII]}$ relation for high-z galaxies of \citet{delooze2014}, while for low metallicity galaxies, brighter [CII] emission would be expected. Overall, this suggests that the AGN is unlikely to be the dominant source powering the [CII] in the Spiderweb galaxy.

Both our APEX [CII] spectrum and the ATCA+VLA CO(1-0) spectrum allowed us to isolate the AGN and CGM components (Fig.~\ref{fig:CO_CII}, Table~\ref{table:obsparams}). Using 
%$L^{\prime}_{\rm co, VLA} \sim (1.9\pm0.6)\times10^{10}\,K\,km\,s^{-1}\,pc^2$
$I_{\rm CO(1-0), VLA}=0.08\pm0.03$ for the AGN \citep{emonts2016}, we found $L_{\rm CO(1-0), VLA}=(9\pm3)\times10^5 L_{\odot}$ and $L_{\rm [CII]}/L_{\rm CO(1-0)} \approx 4200$. This value is  close to the 5200$\pm$1800 found by \citet{gullberg2015} for DSFGs. While this presents a consistent picture where the AGN has a negligible contribution to the cold dust as well as CO(1-0) and [CII] luminosities, it is important to consider the uncertainties in the separation of the AGN and CGM components in one or both of the lines. 
%As we found that also the $L_{\rm [CII]}/L_{\rm FIR}$ is consitent with non-AGN high-redshift galaxies, this suggests that the separation of the AGN contribution to [CII] and/or CO(1-0) may be even uncertain. 
%This similarity may be surprising given that in AGN-dominated sources, the [CII]/CO(1-0) is generally lower than in DSFGs, while the $L_{\rm [CII]}/L_{\rm FIR}$ of the Spiderweb is more consistent with AGN rather than DFSGs. 
We trust the separation in CO(1-0) to be quite reliable as it is based on spatially resolved observations, which vary with distance from the AGN. Moreover, for the compact component near the AGN, \citet{emonts2018} report a thermalized $L^{\prime}_{\rm CO(4-3)}/L^{\prime}_{\rm CO(1-0)} \sim 1$, which is also consistent with AGN excitation. If the spatially isolated AGN component in CO(1-0) were off by a significant amount, this would also affect the CO(4-3) in a similar fashion, which is rather unlikely. 
%On the other hand, we cannot exclude the possibility that some companion sources contribute within the 10\arcsec\ APEX beam, so the AGN component in [CII] (Fig.~\ref{fig:CO_CII}) should rather be seen as an upper limit. 
We therefore conclude that the [CII] luminosity is dominated by the CGM at negative velocities with a possible contribution from the AGN mostly at positive velocities.

Our detection of [CII] in the CGM allows us to better characterize the CGM surrounding one of the most massive high-redshift sources known. \citet{emonts2018} also separated the CGM component in [CI]1-0 and CO(4-3), but those lines likely have more AGN residuals from the central point spread function than for CO(1-0) because the CO emission is thermalized and both [CI] lines are comparatively bright at the central AGN \citep{gullberg2016,emonts2018}.  Conversely, on scales of the CGM, the molecular gas is subthermally excited and has a lower [CI] abundance relative to CO(1-0) compared to the AGN region \citep{emonts2018}, meaning that the fraction of the emission coming from the CGM is larger, and thus easier to separate, in CO(1-0) than in CO(4-3) or [CI]. Our APEX [CII] detection  thus confirms the presence of the CGM component at predominantly negative velocities in CO(1-0), but we lack a signal-to-noise ratio (S/N) to reliably measure the profile at positive velocities. 
To derive the CGM component in CO(1-0), we assumed $L_{\rm CO(1-0), CGM}=L_{\rm CO(1-0), ATCA}-L_{\rm CO(1-0), VLA} = 0.16\pm0.09$\,Jy\,km\,s$^{-1}$ \citep{emonts2016}, yielding $L_{\rm CO(1-0), CGM}=(1.9\pm1.0)\times10^6 L_{\odot}$. This implies $L_{\rm [CII],CGM}/L_{\rm CO(1-0),CGM} \approx 2800$, which for optically thick CO emission suggests low [CII] excitation temperatures, unless the [CII] is also optically thick \citep{gullberg2015}. Normalizing by the FIR luminosity, the CGM component in the Spiderweb galaxy falls in the region of nearby galaxies with average radiation fields $G_0 \sim 10^3$ and densities $n \sim 10^5$\,cm$^{-3}$, assuming the [CII] is mostly dominated by photo-dissociation regions \citep{stacey2010,gullberg2015}.

Although our APEX detection does not provide any spatial information, we predict that the [CII] in the Spiderweb galaxy is likely quite extended because it traces the more extended component in CO(1-0). Moreover, extended [CII] emission has now been regularly observed in several high-redshift objects \citep{cicone2015,fujimoto2020,carniani2020,rybak2020,herrera-camus2021}. Given that the spatial scales can be several tens of kiloparsecs or more, even short-baseline observations at these high frequences  may not be able to detect the full extent of the CGM in [CII] emission. On the other hand, the APEX beam size of 10\arcsec\ corresponds to a physical scale of $\sim$70\,kpc, which is the same as the total extent of the cold molecular gas reservoir in the CGM observed in CO(1-0). 

Our [CII] detection also completes a census of the first four ionization states of carbon. As mentioned earlier, \citet{gullberg2016} and \citet{emonts2018} reported [CI] emission consisting of several spatially and spectrally resolved components, where the AGN component is significantly brighter than the CGM component. The CIII]\,1909\AA\ and CIV\,1549\AA\ lines were first reported in the discovery spectrum of \citet{rottgering1997}, with a line ratio of CIV\,1549\AA\ / CIII]\,1909\AA\ = 0.6$\pm$0.2. Since then, only the CIV\,1549\AA\ line has been observed at a higher S/N and spectral resolution \citep{kurk2003,hatch2008}, suggesting a broad component in the CIV\,1549\AA\ line, but with a total line flux about half of the one reported by \citet{rottgering1997}. We interpret this difference as a combination of slit aperture effects and a low S/N. Only deeper observations with integral field spectrographs can provide more reliable line ratios, but we can nevertheless conclude that the ratio is unlikely to exceed unity. This places the Spiderweb galaxy slightly below average among HzRGs and suggests an AGN photoionization with an ionization parameter log(U)$\approx$-2 \citep{debreuck2000}. We note that \citet{kurk2003} also reported CII\,1334.5\AA\ absorption at $z$=2.1645$\pm$0.0004 with an equivalent width of 2.4$\pm$0.7. This offset of +400\,km\,s$^{-1}$ means that this absorbing gas is part of the CGM surrounding the Spiderweb galaxy, but not associated with the main component we detected in [CII]\,158\,$\mu$m.
Overall,  the integrated carbon budget with bright [CI] and more average [CII], CIII], and CIV argues for a relatively modest excitation with most of the carbon in the lowest ionization state, especially for the AGN component.  Our results suggest that, at least in AGN-dominated regions, carbon excitation to higher levels by an external ionization field is not a major concern when using the [CI] lines as H$_2$ tracers \citep[e.g.][]{glover2016}.

\section{Conclusions}

By extending the tuning range compared to the ALMA Band 9 receivers, we have detected [CII] emission from the Spiderweb galaxy using SEPIA660 on APEX. Our conclusions are:
\begin{itemize} 
\item The [CII] line consists of two velocity components, which by comparing with the CO(1-0) and HeII~1640\AA\ spectra, we identify as being associated with the AGN and the CGM. 
\item The individual components are consistent in terms of CO and FIR luminosity ratios with the AGN and DSFG populations. The CGM component dominates the [CII] flux. 
\item Due to the close correspondence of the CO(1-0) spectrum to the CGM component, we predict that the [CII] flux is extended over several tens of kiloparsecs.
\item The [CII] line completes a census of the first four ionization states of carbon in the Spiderweb galaxy. The atomic [CI] line is most prominent, suggesting relatively low excitation, especially near the AGN.
\end{itemize}

Our detection also illustrates two technical results: first, despite a reduced atmospheric transmission, new science is enabled by extending the frequency range of a Band 9 receiver beyond the original specifications; second, single-dish observations remain essential to detect extended emission from the CGM \citep[see also][]{frayer2018}. However, spending 17 hours in very good weather conditions is at the limit of what a 12m telescope such as APEX can do. Only a large single-dish telescope such as the Atacama Large Aperture Submm Telescope \citep[AtLAST;][]{cicone2019,klaassen2020} will be able to reveal the link between galaxies and their accreting CGM gas streams.

\begin{acknowledgements}
       The results presented in this paper would not have been possible without the dedication of the APEX engineers, operators, astronomers and logistics staff to bring the telescope back online during the covid-19 pandemic. We thank the anonymous referee for their advice that has improved this paper. This publication is based on data acquired with the Atacama Pathfinder Experiment (APEX) under programme ID 0106.A-1003 (ESO). APEX is a collaboration between the Max-Planck-Institut f\"ur Radioastronomie, the European Southern Observatory, and the Onsala Space Observatory. This work has received funding from the European Union’s Horizon 2020 research and innovation programme under grant agreement No 951815. The National Radio Astronomy Observatory is a facility of the National Science Foundation operated under cooperative agreement by Associated Universities, Inc.
HD acknowledges financial support from the Spanish Ministry of Science, Innovation and Universities (MICIU) under the 2014 Ram\'on y Cajal program RYC-2014-15686, from the Agencia Estatal de Investigaci\'on del Ministerio de Ciencia e Innovaci\'on (AEI-MCINN) under grant (La evoluci\'on de los c\'\i umulos de galaxias desde el amanecer hasta el mediod\'\i a c\'osmico) with reference (PID2019-105776GB-I00/DOI:10.13039/501100011033) and acknowledge support from the ACIISI, Consejer\'\i a de Econom\'\i a, Conocimiento y Empleo del Gobierno de Canarias and the European Regional Development Fund (ERDF) under grant with reference PROID2020010107.
\end{acknowledgements}


\end{document}